\documentclass[twocolumn]{aastex63}

\newcommand{\oii}{O\,{\scriptsize II}}
\newcommand{\oiii}{O\,{\scriptsize III}}

\newcommand{\ciii}{C\,{\scriptsize III}}
\newcommand{\sii}{S\,{\scriptsize II}}

\newcommand{\neiii}{Ne\,{\scriptsize III}}

\newcommand{\civ}{C\,{\scriptsize IV}}

\newcommand{\hei}{He\,{\scriptsize I}}
\newcommand{\heii}{He\,{\scriptsize II}}

\newcommand{\lya}{Ly$\alpha$}

\newcommand{\hst}{\textit{HST}}
\newcommand{\jwst}{\textit{JWST}}
\newcommand{\spitzer}{\textit{Spitzer}}
\newcommand{\eazy}{\texttt{EAzY}}

\usepackage{rotating}
\usepackage{float}
\usepackage{amsmath}
\usepackage{graphicx} 
\usepackage[caption=false]{subfig}

\graphicspath{{./}{figures/}}

\shorttitle{}

\begin{document}


\title{The BoRG-\textit{JWST} Survey: Program Overview and First Confirmations of Luminous Reionization-Era Galaxies from Pure-Parallel Observations}

\correspondingauthor{Guido Roberts-Borsani}
\email{guido.roberts-borsani@unige.ch}
\author[0000-0002-4140-1367]{Guido Roberts-Borsani}
\affiliation{Department of Astronomy, University of Geneva, Chemin Pegasi
51, 1290 Versoix, Switzerland}

\author[0000-0002-9921-9218]{Micaela Bagley}
\affiliation{Department of Astronomy, The University of Texas at Austin, 2515 Speedway, Austin, TX, 78712, USA}

\author[0000-0003-2349-9310]{Sof\'ia Rojas-Ruiz}
\affiliation{Department of Physics and Astronomy, University of California, Los Angeles, 430 Portola Plaza, Los Angeles, CA 90095, USA}

\author[0000-0002-8460-0390]{Tommaso Treu}
\affiliation{Department of Physics and Astronomy, University of California, Los Angeles, 430 Portola Plaza, Los Angeles, CA 90095, USA}

\author[0000-0002-8512-1404]{Takahiro Morishita}
\affiliation{IPAC, California Institute of Technology, MC 314-6, 1200 E. California Boulevard, Pasadena, CA 91125, USA}

\author[0000-0001-8519-1130]{Steven L. Finkelstein}
\affiliation{Department of Astronomy, The University of Texas at Austin, 2515 Speedway, Austin, TX, 78712, USA}

\author[0000-0001-9391-305X]{Michele Trenti}
\affiliation{School of Physics, University of Melbourne, Parkville 3010, VIC, Australia}
\affiliation{ARC Centre of Excellence for All Sky Astrophysics in 3 Dimensions (ASTRO 3D), Australia}

\author[0000-0002-7959-8783]{Pablo Arrabal Haro}
\affiliation{NSF’s National Optical-Infrared Astronomy Research Laboratory, 950 N. Cherry Ave., Tucson, AZ 85719, USA}

\author[0000-0002-2931-7824]{Eduardo Ba{\~n}ados}
\affiliation{Max Planck Institut f\"ur Astronomie, K\"onigstuhl 17, D-69117, Heidelberg, Germany}

\author[0000-0003-2332-5505]{\'{O}scar A. Ch\'{a}vez Ortiz}
\altaffiliation{NASA FINESST Fellow}
\affiliation{Department of Astronomy, The University of Texas at Austin, 2515 Speedway, Austin, TX, 78712, USA}

\author[0000-0003-4922-0613]{Katherine Chworowsky}
\altaffiliation{NSF Graduate Fellow}
\affiliation{Department of Astronomy, The University of Texas at Austin, 2515 Speedway, Austin, TX, 78712, USA}

\author[0000-0001-6251-4988]{Taylor A. Hutchison}
\altaffiliation{NASA Postdoctoral Fellow}
\affiliation{Astrophysics Science Division, NASA Goddard Space Flight Center, 8800 Greenbelt Rd, Greenbelt, MD 20771, USA}

\author[0000-0003-2366-8858]{Rebecca L. Larson}
\altaffiliation{NSF Graduate Fellow}
\affiliation{Department of Astronomy, The University of Texas at Austin, 2515 Speedway, Austin, TX, 78712, USA}
\affiliation{School of Physics and Astronomy, Rochester Institute of Technology, Rochester, NY 14623, USA}

\author[0000-0003-4570-3159]{Nicha Leethochawalit}
\affiliation{National Astronomical Research Institute of Thailand (NARIT), Mae Rim, Chiang Mai, 50180, Thailand}

\author[0000-0002-9393-6507]{Gene C. K. Leung}
\affiliation{Department of Astronomy, The University of Texas at Austin, Austin, TX, USA}

\author[0000-0002-3407-1785]{Charlotte Mason}
\affiliation{Cosmic Dawn Center (DAWN)}
\affiliation{Niels Bohr Institute, University of Copenhagen, Jagtvej 128, DK-2200, Copenhagen N, Denmark}

\author[0000-0002-6748-6821]{Rachel S. Somerville}
\affiliation{Center for Computational Astrophysics, Flatiron Institute, 162 5th Avenue, New York, NY 10010, USA}

\author[0000-0001-9935-6047]{Massimo Stiavelli}
\affiliation{Space Telescope Science Institute, 3700 San Martin Drive, Baltimore, MD 21218, USA}

\author[0000-0003-3466-035X]{{L. Y. Aaron} {Yung}}
\affiliation{Space Telescope Science Institute, 3700 San Martin Dr., Baltimore, MD 21218, USA}

\author[0000-0002-3838-8093]{Susan A. Kassin}
\affiliation{Space Telescope Science Institute, 3700 San Martin Drive, Baltimore, MD 21218, USA}
\affiliation{Department of Physics \& Astronomy, Johns Hopkins University, 3400 N. Charles Street, Baltimore, MD 21218, USA}

\author[0000-0001-7840-2972]{Christian Soto}
\affiliation{Space Telescope Science Institute, 3700 San Martin Drive, Baltimore, MD 21218, USA}

\begin{abstract}
We present the BoRG-\textit{JWST} survey, a combination of two \textit{JWST} Cycle 1 programs aimed at obtaining NIRSpec spectroscopy of representative, UV-bright $7<z<10$ galaxy candidates across 22 independent sight lines selected from \textit{Hubble}/WFC3 pure-parallel observations. We confirm the high-$z$ nature of 10 out of 19 observed primary targets through low-resolution prism observations, with the rest revealing themselves unsurprisingly to be $z\sim1-3$ interlopers, brown dwarfs, or yielding inconclusive results. From the MSA observations, we confirm an additional 9 filler sources at $z>5$, highlighting the large abundance of high-redshift galaxies even in individual WFC3 pointings. The primary sample span an absolute magnitude range $-20.4<M_{\rm UV}<-22.4$ mag and harbour UV continuum slopes of $\beta\simeq-2.5$ to $-2.0$, representing some of the most luminous $z>7$ sources currently known and comparable to the brightest sources at $z>10$. Prominent [\oiii]+H$\beta$ lines are found across the full sample, while a stack of sources reveals a plethora of other rest-optical lines and additional rest-UV \ciii]1909 \AA\ emission. Despite their luminosities, none of the low-resolution spectra display evidence for Type 1 AGN activity based on a search for broad-line emission. Lastly, we present a spectroscopic data release of 188 confirmed $0.5\lesssim z\lesssim5.0$ sources from filler MSA observations, highlighting the legacy value of the survey and a representative benchmark for comparisons to deep field observations.
\end{abstract}


\section{Introduction}
The selection of galaxies at high redshift and the characterization of their properties represents a key step towards determining the onset of metal, dust, and structure formation, and the pinpointing of sources that ionized the intergalactic medium (IGM) over the first billion years.

To this end, the revelation and study of an unexpected population of UV-luminous ($M_{\rm UV}\lesssim-21$ mag) sources beyond redshifts of $z\simeq10$ (e.g., \citealt{castellano22,castellano23,finkelstein23,bunker23,curtislake23,carniani24,harikane23,casey23}) by the \textit{James Webb} Space Telescope (\textit{JWST}) has brought our understanding of early galaxy evolution into question, while simultaneously extending studies of galaxy number densities (e.g., \citealt{bouwens23,mason23,donnan24,mcleod24,finkelstein24}), chemical abundances (e.g., \citealt{nakajima23,rb24,deugenio23}), and stellar masses (e.g., \citealt{zavala24,helton24}) out to $z\simeq10-14$.

A thorough understanding of the build up of dark matter halos, the impact of feedback processes, and the efficiency and timescales of star formation governing the evolution of the brightest sources, however, requires accurate determinations of their number densities and properties at both the redshift frontier and at lower redshifts of $6<z<10$ (e.g., \citealt{sanders23,rui24,cameron23}). While much of \textit{JWST}'s focus has been on luminous sources at $z>10$, the decade prior to \textit{JWST} saw remarkable progress in the identification of UV-bright sources at redshifts of $z\simeq6-10$, in large part thanks to the \textit{Hubble} Space Telescope's (\textit{HST}) Wide Field Camera 3 (WFC3) and imaging campaigns with ground-based facilities. The identification of sources beyond $z>6$ with large data sets such as UKIDSS UDS, UltraVISTA, CANDELS, and ZFOURGE (\citealt{lawrence07}, \citealt{mccracken12}, \citealt{grogin11}, \citealt{tilvi13}, respectively), allowed for determinations of the UV luminosity function (UVLF) out to $z\sim10$ (e.g., \citealt{bouwens15,rb16,mcleod16,oesch18,bouwens19,bowler20,finkelstein22}), as well as inferences of their global properties with the aid of the \textit{Spitzer} Space Telescope (e.g., \citealt{labbe13,strait20,rb22}).

However, despite the impressive progress and even with novel capabilities from \textit{JWST}, parameterizations at the brightest end of the UVLF ($M_{\rm UV}<-21$ mag) remain challenging and uncertain. Considering the large volumes required to find them, searches for UV-luminous sources are subject to significant field-to-field variance and confusion \citep{trenti08,robertson2010,willott24}, which can lead to differing parameterizations of the UVLF (e.g., a Schechter function or double power-law; \citealt{bowler15}). Surveys such as CANDELS, CEERS \citep{finkelstein22b,bagley23}, and JADES \citep{eisenstein_jades} benefit from especially deep images that maximise sample purity, however their relatively small and correlated search areas are prone to cosmic variance effects. Conversely, larger fields such as UltraVISTA, COSMOS-Web \citep{casey_cosmosweb}, and Euclid observations \citep{euclid24} benefit from enhanced search areas, but suffer from shallower imaging which can lead to sample contamination by brown dwarfs and $z\sim1-3$ interlopers.

The challenge is exacerbated by the dearth of spectroscopic samples with which to determine pure number counts and the contribution of luminous galaxies to the reionization process. $z>5$ confirmations now number $\sim$1000 sources \citep{rb24,heintz24,issi24,meyer24}, however the vast majority do not probe the most luminous populations (e.g., GNz11, GHZ2, and JADES-GS-z14-0; \citealt{bunker23,castellano24,carniani24}, respectively). Moreover, detections of strong Lyman-$\alpha$ (Ly$\alpha$) emission -- generally attenuated by intervening H\,{\scriptsize I} in a predominantly neutral era -- have remained exclusive to a small number of deep fields suspected to host enlarged ionized bubbles or overdense regions \citep{stark17,tang23,saxena23}. Given those galaxies and other comparatively luminous sources display exceptional intrinsic properties (as traced by extreme rest-frame UV-to-optical features; \citealt{castellano17,endsley21,bunker23,castellano24,kumari24,rb24}) \textit{and} reside in overdense regions of the sky \citep{chen24,endsley22_cosmos,morishita22,tilvi20,larson22,napolitano24}, determining the primary driver of Ly$\alpha$ visibility remains a high priority and requires representative observations over large and independent volumes.

(Pure-)parallel observations represent one promising avenue forward: coordinated parallels or random pointings with a large suite of near-infrared (NIR) filters allow for robust selections of UV-luminous sources from independent sight lines, minimising the effects of galaxy environment and cosmic variance and thus yielding representative samples for unbiased number densities and probes of Ly$\alpha$ opacity. Several pure-parallel \textit{HST} surveys provide excellent opportunities for such selections. Specifically, the Brightest of Reionizing Galaxies \citep[BoRG; PI Trenti, GO programs 11700 and 12572;][]{trenti11,bradley12,schmidt14}, the Hubble Infrared Pure Parallel Imaging Extragalactic Survey \citep[HIPPIES, PI Yan, GO 11702 and 12286;][]{yan2011}, and the WFC3 Infrared Spectroscopic Parallel survey \citep[WISP; PI Malkan, GO 12283, 12902, 13352, 13517, and 14178;][]{atek2010} were each designed with high-redshift galaxy candidate identification in mind. These programs, aimed at high Galactic latitudes to minimize contamination by stellar sources, obtained imaging in multiple ACS and WFC3 filters to constrain F098M, F105M, or F115W dropouts at $z\gtrsim8$, and include hundreds of uncorrelated observations. Recent searches using those data sets have identified a large number of luminous ($M_{\rm UV}\lesssim-21$ mag) galaxy candidates at $7<z<10$ \citep{morishita20,rb22,bagley23}, and ensuing determinations of their $z\sim8-10$ UVLFs \citep{morishita18,rojas20,leethochawalit22} have hinted at enhanced values resembling the evolution of number densities seen at $z>10$ with \textit{JWST}/NIRCam.

Confirmation of those number densities and the role of luminous galaxies in (re)ionizing the universe requires spectroscopic verification, however, which serves as the basis for this paper. Here we introduce the BoRG-\textit{JWST} survey, a multi-facility program aiming to confirm and characterize luminous and representative $z\simeq7-10$ galaxy candidates selected from pure-parallel \textit{HST} imaging with \textit{JWST} spectroscopy. The paper is structured as follows. In Section~\ref{sec:drivers} we present the main scientific drivers of the program, along with a presentation of the sample selection in Section~\ref{sec:sample}. In Section~\ref{sec:confirmations} we present the spectroscopic confirmations resulting from the program, along with their main properties, and conclude the paper with a summary of our findings in Section~\ref{sec:summary}. Throughout this paper we adopt a cosmology with H$_0$=70 km~s$^{-1}$ Mpc$^{-1}$, $\Omega_M$=0.3, $\Omega_\Lambda$=0.7. All magnitudes are quoted in the AB system \citep{oke83}.

\section{Key Science Drivers}
\label{sec:drivers}
The original Brightest of Reionizing Galaxies survey was designed with two key science drivers in mind, namely (i) an unbiased characterization of the bright end of the UVLF, and (ii) providing viable targets for ground-based spectroscopic follow up of rest-frame UV emission lines. The BoRG-\textit{JWST} survey extends and expands on these goals, adding NIRSpec spectroscopic constraints which allow for secure redshifts, simultaneous measurements of \lya\ and rest-frame UV-to-optical spectral features, and the characterization of stellar and interstellar medium (ISM) physics in $z>7$ galaxies.

\begin{figure}
\center
\includegraphics[width=\columnwidth]{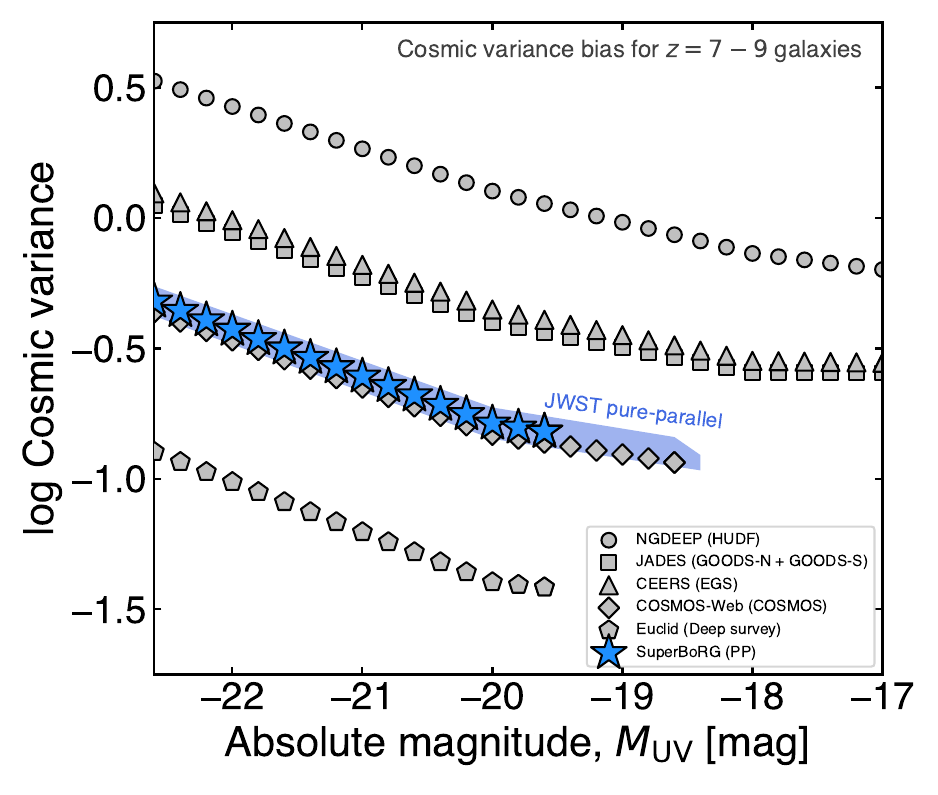}
 \caption{The fractional error due to cosmic variance calculated using the \texttt{CV\_AT\_COSMIC\_DAWN} tool from the \texttt{BlueTides} simulations \citep{bhowmick20}, adopting the $5\sigma$ limiting depths of the largest and deepest \jwst\ imaging surveys (e.g., NGDEEP, JADES, CEERS, and COSMOS-Web; grey symbols; \citealt{bagley24_ngdeep,eisenstein_jades,finkelstein22b,bagley23,casey_cosmosweb}, respectively) compared to the values probed by the independent sight lines of NIRCam pure-parallel surveys (blue shade, including predicted values from PASSAGE, PANORAMIC, and BEACON; GO 1571 PI Malkan, GO 2514 PI Williams, GO 3990 PI Morishita) and the SuperBoRG survey (blue stars; \citealt{morishita21}). Of the legacy fields, only the especially large area of COSMOS-Web is able to match the low cosmic variance probed by SuperBoRG and pure-parallel observations, and even the COSMOS-Web survey represents one contiguous area rather than independent fields. For comparison, the low values from the especially large areas of the Euclid-Deep survey \citep{euclid24} are also plotted.}
 \label{fig:cv}
\end{figure}

Crucially, the BoRG-\textit{JWST} survey provides these constraints over sources selected from a multitude of independent sight lines, minimizing the effects of cosmic variance and offering an alternate benchmark to samples derived from generally reduced or correlated regions of the sky in deep fields. An example of this is shown in Figure~\ref{fig:cv} (calculated using the \texttt{BlueTides} simulations from \citealt{bhowmick20}, with predicted survey areas and $5\sigma$ depths), where we illustrate the fractional error due to cosmic variance for $z=7-9$ galaxies derived for a number of \textit{JWST} surveys. 
The smallest areas introduce the largest variance, and only the largest (but contiguous) surveys
are able to compete with the low values probed by independent BoRG pointings and NIRCam pure-parallel surveys. For comparison, we also show results for the $\sim$50 deg$^{2}$ probed by the Euclid-Deep survey, which provide by far the lowest values.
With this in mind, we list and describe the primary (and expanded) science goals of BoRG-\textit{JWST} here below, which will be addressed in a number of future papers.

\subsection{Constraining the UV Luminosity Function: How Prevalent Are Luminous Sources?}
\begin{figure*}
\center
\includegraphics[width=\textwidth]{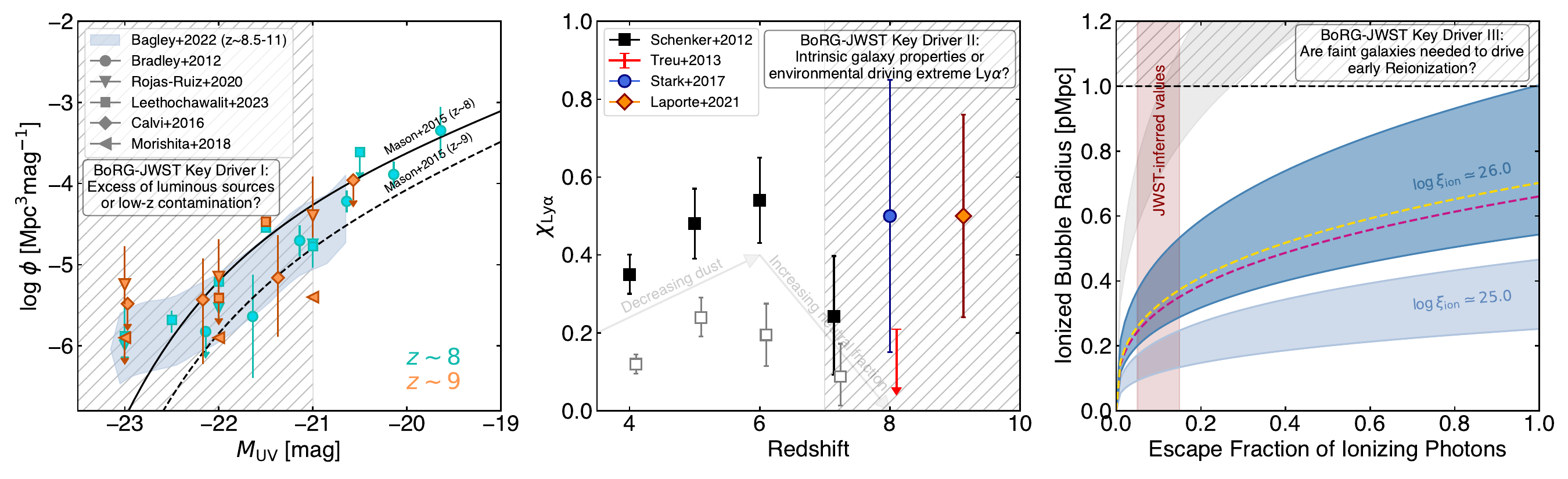}
 \caption{A schematic representation of the key science drivers for the BoRG-\textit{JWST} survey, namely: (i) constraining the bright end of the $7<z<10$ UV luminosity function (left panel), (ii) determining the opacity of Ly$\alpha$ in UV-luminous galaxies (middle panel), and (iii) constraining the contributions of UV-luminous sources to the ionizing UV photon budget within the first billion years (right panel). In each case, the hatched region points to the region of relevant parameter space BoRG-\textit{JWST} aims to probe.}
 \label{fig:goals}
\end{figure*}

The majority of pre-\textit{JWST} studies
agreed on a rapid evolution at the bright end of the UVLF between $z\simeq4-10$ \citep{bouwens15,bowler15,finkelstein15}, however the parametrization of those number counts remained debated. While some studies favoured a rapid decline in galaxies characterized by a Schechter function, others preferred a double power-law (DPL) signalling a combination of enhanced number densities, a lack of AGN feedback, and/or a lack of dust obscuration. This is particularly true at redshifts beyond $z\simeq8-9$, where field-to-field variance from small areas of the sky (introducing $\sim25$\% uncertainty on galaxy number counts in deep fields; \citealt{trenti11}) often lead to contrasting results, and translations to a cosmic star formation rate density (SFRD) illuminated a tension in the literature: Schechter parametrizations generally yielded a rapid and decreasing evolution in line with expectations from the evolution of the dark matter halo mass function \citep[e.g.,][]{oesch18,bouwens21,harikane22}, while DPL parametrizations favoured a more gradual evolution extending towards earlier times \citep[e.g.,][]{mcleod16,bowler20,finkelstein22}.
The uncorrelated sight lines of BoRG, HIPPIES, and WISP represented an opportunity for an independent assessment at the bright end of the UVLF, without the cosmic variance characteristic of deep field inferences. With both \hst/ACS and WFC3 imaging allowing for the selection of $z>7$ ``Lyman-break'' dropout sources, multiple BoRG studies have evaluated the number densities and parametrizations of the UVLF primarily from $z\sim8$ to $z\sim9$ \citep{trenti11,bradley12,schmidt14,calvi16,livermore18,morishita18,rojas20,leethochawalit22,rb22,bagley23}. Those works are summarized in the left panel of Figure~\ref{fig:goals}, where we plot the number densities as a function of absolute magnitude for BoRG studies, where available. Good agreement was found with theoretical predictions at $z\sim8$ \citep{mason15}, while the scatter at $z\sim9$ made interpretations challenging. Notably, an apparent excess of sources at the most luminous end was found, however the degree to which those reflect enhanced number densities of high-$z$ galaxies (as now seen by \textit{JWST}/NIRCam) or contamination from low-$z$ interlopers remains an open question and the first key driver of the BoRG-\textit{JWST} survey.

\subsection{Determining the Opacity of Ly$\alpha$ from Independent Lines of Sight}
The visibility of \lya\ in Lyman-break galaxies has long been used as a tracer for the evolving neutral fraction of the intergalactic medium over the course of the first billion years of the universe \citep{fontana10,schenker12,treu13,schenker14,pentericci14,mason19}. Spectroscopic follow up of magnitude-limited samples with ground-based facilities typically yield(ed) very few, if any, convincing detections above $z\simeq6$ \citep[e.g.,][]{pentericci11,finkelstein13,jung19}. As a remarkable exception to this rule, the follow up of especially luminous candidates harbouring intense [\oiii]+$H\beta$ line emission yielded significantly higher detection rates of \lya\ \citep{oesch15,zitrin15,rb16,stark17,endsley21,tilvi20,endsley22_cosmos,larson22}, challenging the utility of the line as a tracer for cosmic reionization and demonstrating the patchiness of the process. The picture is summarized in the middle panel of Figure~\ref{fig:goals}.

However, the vast majority of those detections lie in the CANDELS-EGS, GOODS, and COSMOS fields, suspected to host enlarged ionized bubbles and serving as a dramatic illustration of the effects of cosmic variance. The secure redshifts and wide wavelength coverage afforded by \jwst/NIRSpec have now opened a new window for \lya\ studies, allowing for $z>6$ constraints on the line independently of galaxy properties and corroborating pre-\jwst\ claims \citep[e.g.,][]{tang23,bunker23,saxena23,rb23,larson23,witten24}.
A close look at their average spectra has revealed significant differences in the intrinsic properties of \lya-emitters compared to their \lya-attenuated counterparts \citep{rb24} and the overdense environments around many UV-luminous \lya-emitters confirmed \citep{witstok24}.
Thus, linking \lya\ detection rates to intrinsic galaxy properties, overdense environments, or large-scale ionized bubbles remains challenging and necessitates independent sight lines. The second key driver of the BoRG-\textit{JWST} survey is therefore to disentangle the effects of the IGM, the effects of galaxy clustering and environment, and the intrinsic properties of the host galaxies in regulating the visibility of strong \lya.


\subsection{The Ionizing Capabilities of Luminous Sources and Contributions to reionization}
Constraining the relative contributions of faint and luminous sources to the (re)ionization of the intergalactic medium had, until recently, relied on photometric number counts of $z\gtrsim6$ sources and uncertain assumptions on the output of their ionizing UV photons (i.e., the Lyman-continuum photon production efficiency and the escape fraction of ionizing UV photons). While the predominant view has been that faint galaxies dominated the ionizing budget at early times \citep{schmidt14,robertson15,yung20a,yung20b,finkelstein19}, the surprising revelation of \lya\ emission in $z\simeq7-9$ UV-luminous ($M_{\rm UV}\simeq-21$ to $-22$ mag) sources from ground-based spectroscopy \citep{finkelstein15,zitrin15,oesch15,stark17,tilvi20,larson22} brought about suggestions of a rapid and late reionization process \citep{naidu20}, one where luminous sources with exceptional ionizing capabilities could carve out early ionized bubbles \citep{mason18,rb23,larson22}. In a simplistic framework (i.e., neglecting the effects of large-scale ionized regions or emission line velocity offsets), much of this picture relies on luminous sources harbouring sufficiently high ionizing photon production efficiencies, $\xi_{\rm ion}$, and/or escape fractions, $f_{\rm esc}$, for those ionized bubbles to become large enough that \lya\ is shifted out of resonance (see examples by \citealt{endsley21} and \citealt{larson22}).

The scenario is summarized in the right panel of Figure~\ref{fig:goals}, where we use the formalism adopted by \citet{endsley21} and \citet{larson22} to determine the ionized bubble radius created by UV-luminous sources, based on assumptions of $\xi_{\rm ion}$ for a range of $f_{\rm esc}$ values. The degeneracy between the two parameters clearly illustrates the need for empirical constraints on the ionizing capabilities of luminous sources, and although medium-band \textit{JWST} imaging is now making impressive progress in this regard \citep{endsley23,simmonds24}, direct spectroscopic constraints on the most luminous sources remain scarce.
As such, the third and final key driver of the BoRG-\textit{JWST} survey is to determine the ionizing photon production efficiencies of luminous sources and assess their relative contributions to the reionization process of the IGM.

\section{Photometric Selections and NIRSpec Data Set}
\label{sec:sample}
The sample of $z\simeq7-10$ sources selected for the \textit{JWST}/NIRSpec follow up presented here derive from the catalogs of photometrically-selected sources presented by \citet{rb22}, \citet{rojas20}, and \citet{bagley24}. We provide a brief summary of those works here, and the selection of those sources for NIRSpec follow up.

\subsection{GO 1747}
The GO 1747 program (PI Roberts-Borsani, 25.1 hrs) was designed to follow up 10 of the most luminous and robust candidates from the sample of \textit{HST}-selected $z\simeq7-10$ sources by \citet{rb22}. The parent sample of sources was constructed using 288 random sight lines (approximately $\sim$1267 arcmin$^{2}$) of pure-parallel imaging data \citep{morishita21}.
As a first step, high-$z$ candidates detected in either an F125W or F140W+F160W image were selected based on their NIR colors, targeting Lyman-break dropouts with the use of ACS and WFC3 images. The selections were split into three criteria, each tailored to a specific redshift range and the available filters: $z\simeq8$ dropouts were selected using the F098M and/or F105W filters, $z\simeq9$ dropouts were selected with the F105W filter, and $z\simeq10$ dropouts were selected using the F125W filter. All selections adopted additional S/N cuts for non-detections blueward of the break and detections in NIR filters redward of it.
Additionally, all sources with a \texttt{SExtractor} stellarity parameter greater than 0.95 were excluded from the final samples to avoid confusion with nearby stars, while a second color cut using \textit{Spitzer}/IRAC imaging (where available) of $H_{\rm 160}-[3.6]<1.4$ was applied to reduce potential contamination from red $z\sim2-4$ galaxies.

Secondly, in order to refine the photometric redshifts of the remaining sources and discard any objects suspected to be brown dwarf or low redshift contaminants, each object was fit with the photometric redshift fitting code, \texttt{EAzY} \citep{brammer08}, adopting a skew-normal prior with galaxy templates to account for the relative abundances of $z\simeq2-4$ sources compared to high-$z$ galaxies resulting from the aforementioned color cuts, or brown dwarf templates from the SpeX Prism Library \citep{burgasser14}. In total, the procedure resulted in 32 galaxy candidates at $z\gtrsim8$ covered by 29 independent sight lines.

\begin{figure*}
\center
\includegraphics[width=\textwidth]{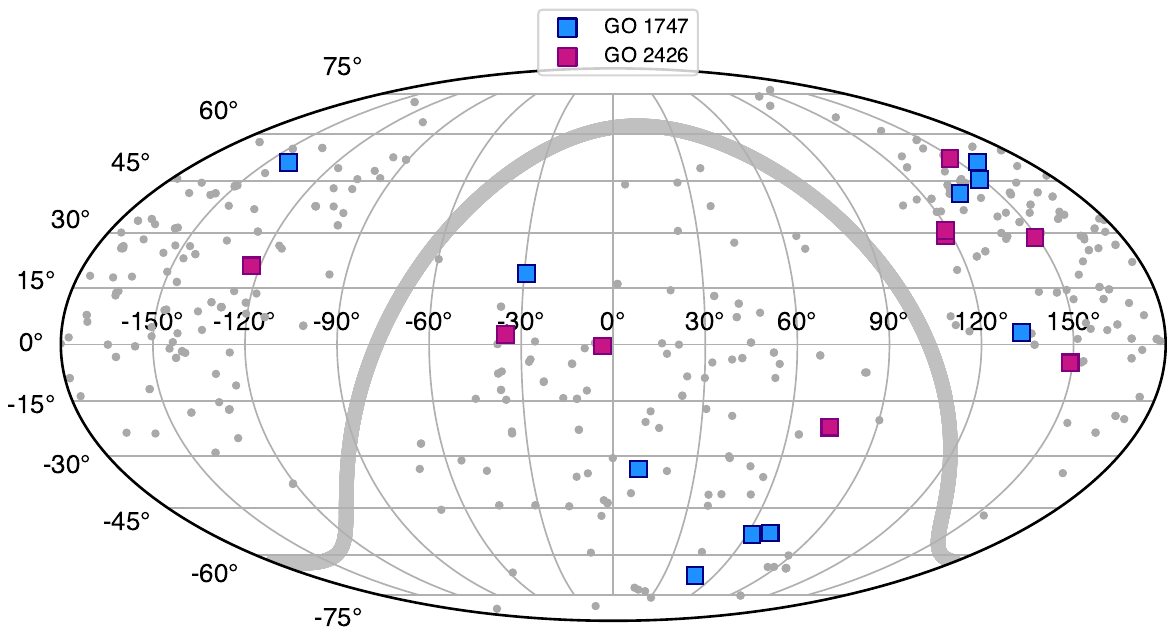}
 \caption{The positions of all 316 SuperBoRG pointings and 45 WISP pointings (grey points; \citet{atek2010,morishita21}), in ecliptic coordinates. Fields targeted for follow-up spectroscopy with \textit{JWST}/NIRSpec are illustrated as color squares (blue for GO 1747 and purple for GO 2426). The independent and uncorrelated nature of the pointings make pure-parallel selections of luminous, high-redshift sources ideal for the construction of samples with minimal cosmic variance.}
 \label{fig:pointings}
\end{figure*}

For the NIRSpec selection, priority was given to sources with \textit{Spitzer}/IRAC coverage to minimize chances of low-$z$ contamination, as well as those with multiple bands blueward of the Lyman-$\alpha$ break. In particular, the highest priority was given to two sources in the catalog with existing Keck/MOSFIRE observations, taken by \citet{morishita20} and \citet{rb22} in search of Ly$\alpha$ emission (0853$+$0310\_112 and 2203$+$1851\_1071 in the latter's catalog, respectively). Furthermore, one field (0859$+$4114) contained two high-$z$ candidates from the fiducial sample of \citet{rb22}, and was thus included to maximise the number of high-$z$ targets (i.e., 11 targets over 10 unique pointings).
While the primary focus of the program was the confirmation and characterization of those high-$z$ sources, the observations were carried out in the Multi Shutter Array (MSA) mode, for a number of reasons. The first was to maximise the success of the target acquisition, given the tendency of BoRG pointings to be carried out at high galactic latitudes where bright stars are scarce: most such observations lacked nearby stellar sources to perform a blind offset to the high-$z$ of interest, and thus we opted for the MSA target acquisition (MSATA) mode where at least five sources (stars, compact galaxies, etc) were required to lie in the MSA for target acquisition. The second was to maximise ancillary science and community data products by filling the rest of the MSA with a mixture of $z>5$ photometric candidates not included in the parent sample, $z\simeq1-3$ star-forming galaxies, and $z\simeq2-4$ passive galaxy candidates.
The total allocated time of 25.5 hours was split approximately equally between fields, amounting to an exposure time of $\sim$2699 seconds per source (4084-4449 seconds for two fields in the sample). Observations were carried out with a NRSIRS2RAPID readout pattern, together with 36-60 groups/integration and a five shutter slitlet, allowing for local background subtraction using adjacent shutters. The decision to opt for a five-shutter rather than a three-shutter slitlet was motivated primarily to reduce possible self-subtraction of sources -- which generally extend beyond the 0.2$\times$0.46$''$ shutter size -- in the background subtraction process, given the use of adjacent shutters in the MSA, as well as for contingency in the event of failed or contaminated observations. The fields were observed between March 21 and December 1 2023, with the exception of 0853+0309 due to a failed target acquisition. An illustration of the compiled SuperBoRG pointings, the high-$z$ selections of \citet{rb22}, and the sources with \textit{JWST}/NIRSpec follow up is shown in Figure~\ref{fig:pointings}.

\subsection{GO 2426}
The GO 2426 program (Co-PIs Bagley and Rojas-Ruiz, 18.2 hrs) was designed as a spectroscopic follow-up of two additional complementary searches for $z\gtrsim8$ galaxy candidates in \hst\ pure-parallel observations. 
\citet{rojas20} explored 90 pointings from the BoRG survey, focusing on observations that included the F140W filter for improved selection of $z>9$ galaxies. \citet{bagley24} considered the remaining fields from BoRG and HIPPIES,
and also included 45 fields from the WISP survey.
Together, these two papers selected candidate high-redshift galaxies from 224 uncorrelated, independent pointings covering $\sim$1040 arcmin$^2$. While the collection of fields have a variety of filter coverages and depths, all include deep ($5\sigma>26.5$) F160W imaging, additional near-IR imaging with a subset of F098M, F105W, F110W, F125W and F140W, and optical imaging using at least one of F350LP, F600LP, F606W, and F814W. Many fields were also observed with \spitzer/IRAC 3.6$\mu$m and 4.5$\mu$m. 

The two studies created F160W-selected catalogs in each field, measured empirical photometric uncertainties using a location- and aperture-size-dependent noise calculation, and modelled IRAC source fluxes with \texttt{Galfit} \citep{peng2010} to deblend them from neighboring sources. Photometric redshifts were computed using \eazy\ with a flat luminosity prior, chosen 
to avoid biasing the selection against true high-redshift galaxies, but at the expense of potentially higher contamination from lower-redshift sources.  
Candidate $z \sim 8-10$ galaxies were selected using a requirement on the detection significance ($5\sigma$) in F160W and/or F140W where available; cuts on the S/N of detections in any available ACS and WFC3 optical bands; a magnitude cut at the bright end ($m_{\mathrm{F160W}} > 22$) to avoid stars; and a lower limit on the measured half light radius to remove cosmic rays and hot pixels in the undithered pure-parallel data. Sample selection was based on the shape of the photometric redshift probability distribution function calculated for each source, selecting sources with a majority of probability at high redshift. Both studies included a comparison of candidate galaxy colors with those of low-mass stars and brown dwarfs, and performed a detailed visual inspection of all candidates.

The procedures resulted in a final sample of 18 candidate $z=9-10$ galaxies with $m_{F160W}<26.5$. For sources selected for NIRSpec follow up, the sample was restricted to the 11 sources with \spitzer/IRAC coverage in at least one filter, as measurements at the longer wavelengths have the power to discern between the flat SED of a true high-redshift galaxy, the redder color of a lower-redshift dusty galaxy, and even cool stars that can contaminate Lyman-$\alpha$ break samples. The sample includes one target from a WISP observation, Par335\_251, representing the first $z>8$ candidate from the survey. It also includes two pairs of candidates in the same BoRG pointings, (Par0953+5153\_1777 and Par0953+5153\_1655) and (Par0956+2847\_169 and Par0956+2847\_1130). While these sources are likely too far apart to reside in the same ionized bubble, they support the idea that our sample of bright, massive galaxies may trace rare, early cosmic overdensities \citep[as e.g.,][]{castellano2016,castellano2018}. 

The main goal of GO 2426 was to spectroscopically confirm the candidates through the detection of the \lya\ break. Using the ETC, we created simulated spectra of the best-fitting \eazy\ high-redshift and low-redshift templates for each candidate. We determined the exposure specification needed to achieve a S/N$\sim$3 per pixel in the continuum, which would allow us to reliably distinguish between the high and low redshift solutions (a spectral break versus the shallower slope of a red and/or dusty galaxy at $z\sim2-3$) even in the absence of any emission lines. We opted to use the S400A1 fixed slit, as several of the pure-parallel pointings lacked sufficient \textit{HST} imaging to allow us to adequately fill the MSA, and the wider 0\farcs4 slit is more forgiving of centroiding uncertainties and less affected by slit losses. We used the NRSIRS2RAPID readout pattern with 12 groups per integration and 8 total dithers -- a 2-point primary dither, which placed the targets at two positions along the slit $\sim$1\farcs6 apart, and sub-pixel dithers in both the spatial and spectral directions. These specifications were designed to allow us to cleanly background-subtract the spectra, remove detector effects, and improve the sub-pixel sampling. Target acquisition was performed with the Wide Aperture Target Acquisition (WATA) mode. Eight of the eleven observations used an offset star, and the remaining three acquired directly on target because a suitable point source (isolated and unsaturated in our \textit{HST} imaging) was not available within the visit splitting distances for the observations. Nine targets were observed between March 12, 2023, and January 7, 2024, with the remaining two observations planned for October-December 2024.

The 2426 target acquisition and observations were successful with the following two exceptions. First, the target Par0956-0450\_684 had a very close neighbor that would have fallen inside the WATA aperture during acquisition, and so this observation's target acquisition required a creative approach. We acquired on the target using initial coordinates offset by $\sim$0\farcs3,  such that the target would be alone in the WATA window. However, this high-risk high-reward maneuver was not successful, and the target did not end up in the slit. Second, the WISP target (Par335-251) resulted in a non-detection. The acquisition images for this observation are empty, and so it is unclear whether the acquisition failed in some way possibly placing the target outside the slit. Finally, following our first several observations, we replaced the lowest-confidence target from GO 2426 (a single-band detection and therefore at a higher risk of being a spurious source in the \textit{HST} imaging) with 0853+0309 from GO 1747. Target 0853+0309 was originally skipped in GO 1747 due to a failed acquisition, but is now one of the two planned GO 2426 observations for Fall 2024.

\section{NIRSpec Prism Spectroscopy}
\label{sec:confirmations}
\subsection{Data Reduction}
We reduce the NIRSpec spectra according to the procedure outlined in \citet{rb24}, which uses a mixture of official STScI pipeline routines, customes codes, and the latest available reference files from STScI at the time of writing (i.e., \texttt{jwst\_1180.pmap} or later). We begin by downloading the uncalibrated files from the MAST directory, and convert these to count-rate files via the \texttt{Detector1Pipeline} routine. However, prior to the ramp-fitting and gain-scaling steps, we utilize an additional routine\footnote{\url{https://github.com/mpi-astronomy/snowblind}} to mask snowballs and cosmic rays, allowing for a better recovery of pixels in the countrate images than would have been allowed if applied over the rate files. The resulting countrate images are then passed through the \texttt{preprocessing} step of \texttt{msaexp}, which applies a number of STScI pipeline routines and custom codes to correct for flat-fielding, 1/$f$ removal, bias removal, and other detector artefacts, prior to the extraction of 2D spectra from individual MSA shutters or the s400a1 fixed slit. Still using \texttt{msaexp}, the extracted spectra are then drizzled to a common reference frame, prior to being used for a local background subtraction along the adjacent shutters of their pseudo-slitlet (or dither positions, in the case of the fixed slit), and combined into an inverse-variance weighted median image (with outlier rejection). The 1D spectrum is extracted and combined from the individual, background-subtracted 2D exposures using a Gaussian fit to the galaxy spectrum along the spatial axis, at the expected position of the trace.
For each spectrum, the spectroscopic redshift is then derived using the redshift-fitting module of \texttt{msaexp}, which utilizes a series of galaxy and line templates from \texttt{EAzY} to fit the entire wavelength range covered by the prism, including features such as the Lyman-$\alpha$ break, continuum breaks, and emission line features.

\subsection{First Confirmations of EoR Galaxies from Pure-Parallel Observations}
In total, we report 14 sources with secure $z\geqslant5$ redshifts from the MSA observations of GO 1747, and 4 sources at $z\sim8$ from the fixed slit observations of GO 2426. Of the primary sources observed between the two programs (at the time of writing), 6/10 from GO 1747 and 4/9 from GO 2426 were confirmed as high-$z$ galaxies, while the rest resulted in either (i) $z\sim2-3$ contaminants (3/10 and 3/9 from GO 1747 and GO 2426, respectively), (ii) a brown dwarf (1/10 from GO 1747), or (iii) yielded inconclusive observations (2/9 from GO 2426). A comparison of the brown dwarf spectrum to the SpeX Prism Libraries of brown dwarf spectra suggests a close match to a $T2$ dwarf. Additionally, one primary source (par0956$+$2847\_169) revealed itself to be part of a multi-component system, for which a clear secondary trace was revealed in the 2D spectrum (dubbed par0956$+$2847\_169-neigh) of GO 2426 and which likely represents a merging system at $z\sim8$. The source(s) and the implications for number counts in the UVLF will be discussed by Rojas-Ruiz et al. (in prep). We report the spectroscopic redshifts and a number of spectro-photometric properties for the full high-$z$ sample in the top half of Table~\ref{tab:confirmations}.

\begin{deluxetable*}{llllccccc}
\tabletypesize{\footnotesize}
\tablewidth{0pt}
\tablecaption{Spectroscopic redshifts and a number of additional properties for confirmed $z\geqslant5$ sources in the BoRG-\textit{JWST} program. Primary targets are marked with a dagger symbol.}
\label{tab:confirmations}
\tablehead{
\colhead{NIRSpec Name} &
\colhead{Field} &
\colhead{RA} & 
\colhead{Dec} &
\colhead{PID} &
\colhead{$z_{\rm spec}$} &
\colhead{$H_{\rm 160}$} &
\colhead{$M_{\rm UV}$} &
\colhead{$\beta$}
}
\startdata
\\
\multicolumn{9}{c}{Secure Sources}\\[0.1cm]
\hline
par0956$+$2847\_1130$^{\dag}$ & 0956$+$2847 & 149.12277 & 28.79200 & GO 2624 & 8.490 & 26.4$\pm$0.2 & $-$20.69$\pm$0.22 & $-$2.37$\pm$0.36 \\
par0953$+$5153\_1777$^{\dag}$ & 0953$+$5153 & 148.29483 & 51.87519 & GO 2624 & 8.440 & 26.7$\pm$0.5 & $-$20.44$\pm$0.41 & $-$2.23$\pm$0.26 \\
1747\_199$^{\dag}$ & 1033$+$5051 & 158.18652 & 50.84159 & GO 1747 & 8.316 & 26.0$\pm$0.2 & $-$21.28$\pm$0.17 & $-$2.21$\pm$0.06 \\
par0956$+$2847\_169$^{\dag}$ & 0956$+$2847 & 149.11331 & 28.81224 & GO 2624 & 8.230 & 24.9$\pm$0.1 & $-$22.34$\pm$0.06 & $-$2.02$\pm$0.10 \\
1747\_732$^{\dag}$ & 0440$-$5244 & 69.94617 & -52.73181 & GO 1747 & 8.226 & 25.8$\pm$0.2 & $-$21.47$\pm$0.15 & $-$2.21$\pm$0.06 \\
par0956$+$2847\_169-neigh & 0956$+$2847 & 149.11351 & 28.81228 & GO 2624 & 8.205 & 25.6$\pm$0.1 & $-$21.45$\pm$0.09 & $-$2.50$\pm$0.17 \\
par0953$+$5153\_1655$^{\dag}$ & 0953$+$5153 & 148.29471 & 51.87709 & GO 2624 & 8.030 & 26.5$\pm$0.2$^{*}$ & $-$20.68$\pm$0.19$^{*}$ & $-$2.05$\pm$0.13 \\
1747\_902$^{\dag}$ & 0955$+$4528 & 148.82808 & 45.48953 & GO 1747 & 7.905 & 25.7$\pm$0.2 & $-$21.10$\pm$0.19 & $-$2.15$\pm$0.09 \\
1747\_1081$^{\dag}$ & 2203$+$1851 & 330.69284 & 18.85811 & GO 1747 & 7.838 & 25.7$\pm$0.2 & $-$21.59$\pm$0.14 & $-$1.63$\pm$0.06 \\
1747\_817$^{\dag}$ & 0409$-$5317 & 62.33714 & -53.25899 & GO 1747 & 7.556 & 26.6$\pm$0.1 & $-$20.74$\pm$0.13 & $-$2.42$\pm$0.09 \\
1747\_1425$^{\dag}$ & 1437$+$5044 & 219.21054 & 50.72599 & GO 1747 & 7.553 & 25.7$\pm$0.1 & $-$21.38$\pm$0.06 & $-$2.30$\pm$0.07 \\
1747\_138 & 0859$+$4114 & 134.82222 & 41.22362 & GO 1747 & 7.179 & 25.5$\pm$0.2 & $-$21.42$\pm$0.16 & $-$2.52$\pm$0.05 \\
1747\_m5 & 0314$-$6712 & 48.42623 & -67.21561 & GO 1747 & 6.525 & 25.4$\pm$0.1 & $-$21.38$\pm$0.06 & $-$1.51$\pm$0.01 \\
1747\_587 & 0314$-$6712 & 48.42864 & -67.21155 & GO 1747 & 6.512 & 27.9$\pm$0.5 & $-$19.15$\pm$0.33 & $-$2.23$\pm$0.47 \\
1747\_1084 & 0037$-$3337 & 9.27036 & -33.61737 & GO 1747 & 6.305 & 26.7$\pm$0.2 & $-$20.20$\pm$0.19 & $-$2.21$\pm$0.12 \\
1747\_269 & 0037$-$3337 & 9.29223 & -33.63230 & GO 1747 & 6.290 & 27.0$\pm$0.4 & $-$20.29$\pm$0.26 & -- \\
1747\_528 & 0314$-$6712 & 48.41869 & -67.21225 & GO 1747 & 6.021 & 26.5$\pm$0.1 & $-$20.31$\pm$0.11 & $-$1.60$\pm$0.01 \\
1747\_1257 & 0314$-$6712 & 48.37564 & -67.20327 & GO 1747 & 5.891 & 27.7$\pm$0.4 & $-$19.69$\pm$0.26 & $-$2.45$\pm$0.17 \\
1747\_412 & 2203$+$1851 & 330.72638 & 18.84254 & GO 1747 & 5.755 & 24.9$\pm$0.1 & $-$21.98$\pm$0.10 & $-$2.21$\pm$0.02 \\
\\
\multicolumn{9}{c}{Possible confirmations}\\[0.1cm]
\hline
1747\_449 & 0314$-$6712 & 48.45539 & -67.21381 & GO 1747 & 9.744 & 27.0$\pm$0.2 & $-$19.5$\pm$0.2 & -99.0$\pm$-99.0 \\
1747\_1050 & 1437$+$5044 & 219.17657 & 50.72104 & GO 1747 & 8.166 & 27.1$\pm$0.2 & $-$19.1$\pm$0.2 & -99.0$\pm$-99.0 \\
1747\_18 & 1033$+$5051 & 158.19745 & 50.83707 & GO 1747 & 7.984 & 25.5$\pm$0.2 & $-$20.8$\pm$0.2 & -99.0$\pm$-99.0 \\
1747\_209 & 1437$+$5044 & 219.22220 & 50.70808 & GO 1747 & 7.934 & 26.7$\pm$0.2 & $-$19.6$\pm$0.2 & -99.0$\pm$-99.0 \\
1747\_1059 & 2203$+$1851 & 330.71237 & 18.85517 & GO 1747 & 7.902 & 25.1$\pm$0.1 & $-$21.1$\pm$0.1 & -99.0$\pm$-99.0 \\
1747\_525 & 0859$+$4114 & 134.81369 & 41.23435 & GO 1747 & 7.572 & 25.7$\pm$0.2 & $-$20.4$\pm$0.2 & -99.0$\pm$-99.0 \\
1747\_m1 & 1437$+$5044 & 219.17234 & 50.72016 & GO 1747 & 7.222 & 27.7$\pm$0.3 & $-$18.4$\pm$0.3 & -99.0$\pm$-99.0 \\
1747\_1406 & 1437$+$5044 & 219.21185 & 50.72567 & GO 1747 & 7.216 & 27.2$\pm$0.3 & $-$18.9$\pm$0.3 & -99.0$\pm$-99.0 \\
1747\_m9 & 0314$-$6712 & 48.46789 & -67.21606 & GO 1747 & 7.096 & 27.4$\pm$0.3 & $-$18.7$\pm$0.3 & -99.0$\pm$-99.0 \\
1747\_10001 & 1437$+$5044 & 219.22397 & 50.72599 & GO 1747 & 6.884 & 26.4$\pm$0.1 & $-$19.7$\pm$0.1 & -99.0$\pm$-99.0 \\
1747\_1288 & 1437$+$5044 & 219.23088 & 50.72403 & GO 1747 & 6.851 & 27.4$\pm$0.6 & $-$18.7$\pm$0.6 & -99.0$\pm$-99.0 \\
1747\_196 & 0859$+$4114 & 134.84496 & 41.22648 & GO 1747 & 6.417 & 25.9$\pm$0.3 & $-$20.1$\pm$0.3 & -99.0$\pm$-99.0 \\
1747\_847 & 0314$-$6712 & 48.37633 & -67.20784 & GO 1747 & 6.212 & 26.8$\pm$0.2 & $-$19.2$\pm$0.2 & -99.0$\pm$-99.0 \\
1747\_1372 & 2203$+$1851 & 330.70947 & 18.86169 & GO 1747 & 6.135 & 25.6$\pm$0.1 & $-$20.3$\pm$0.1 & -99.0$\pm$-99.0 \\
1747\_954 & 0314$-$6712 & 48.48138 & -67.20667 & GO 1747 & 5.992 & 27.0$\pm$0.2 & $-$18.9$\pm$0.2 & -99.0$\pm$-99.0 \\
\enddata
\tablenotetext{*}{Corrected for gravitational lensing, assuming $\mu=2.22$.}
\end{deluxetable*}

Moreover, we highlight the spectrum of the first confirmation of this program, a luminous $z\sim8.3$ source, in Figure~\ref{fig:spectrum}, and show the rest of the confirmed spectra in the Appendix. For completeness, we also show in the Appendix the spectra of the primary targets which revealed themselves to be low-$z$ interlopers. The MSA observations also yielded a number of potential high-$z$ objects for which the redshift could not be confidently established due to low S/N spectral features which could not be unambiguously determined.
Those sources, therefore, are listed in the bottom half of Table~\ref{tab:confirmations} along with their high-$z$ solution. 

\begin{figure*}
\center
\includegraphics[width=\textwidth]{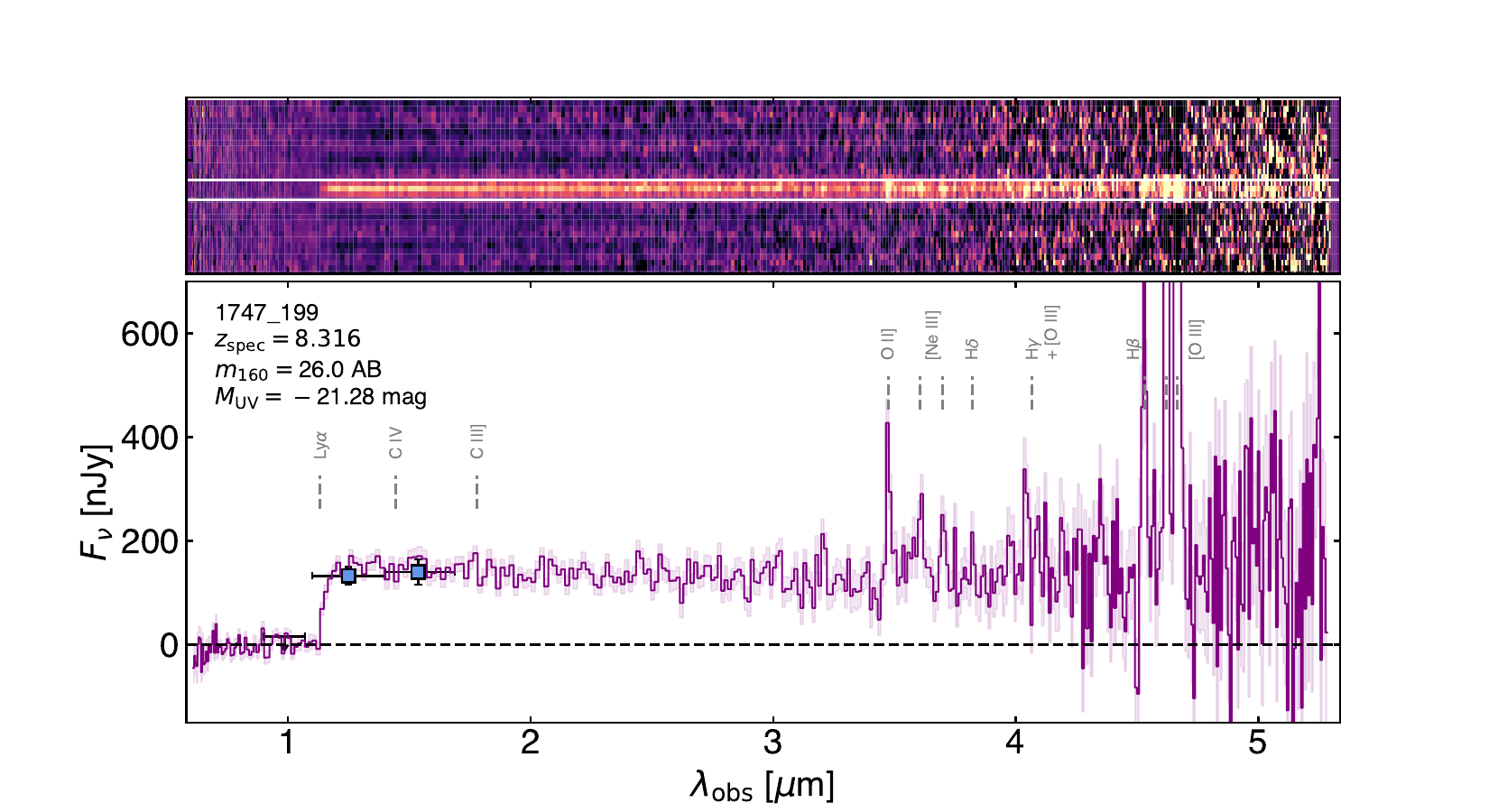}
 \caption{A highlighted BoRG-\textit{JWST} prism spectrum (purple line with pink-shaded $1\sigma$ uncertainties) of 1747\_199, the first confirmation of the program at $z\simeq8.3$, and one of the most luminous sources at $M_{\rm UV}=-21.3$ mag. A strong Lyman-$\alpha$ is found, along with clear rest-frame UV and optical continuum, and a plethora of rest-frame optical emission lines (indicated by grey dashed lines). For comparison, we also plot the \hst\ photometry used in its selection (blue squares and black upper limits.)}
 \label{fig:spectrum}
\end{figure*}

\begin{figure}
\center
\includegraphics[width=\columnwidth]{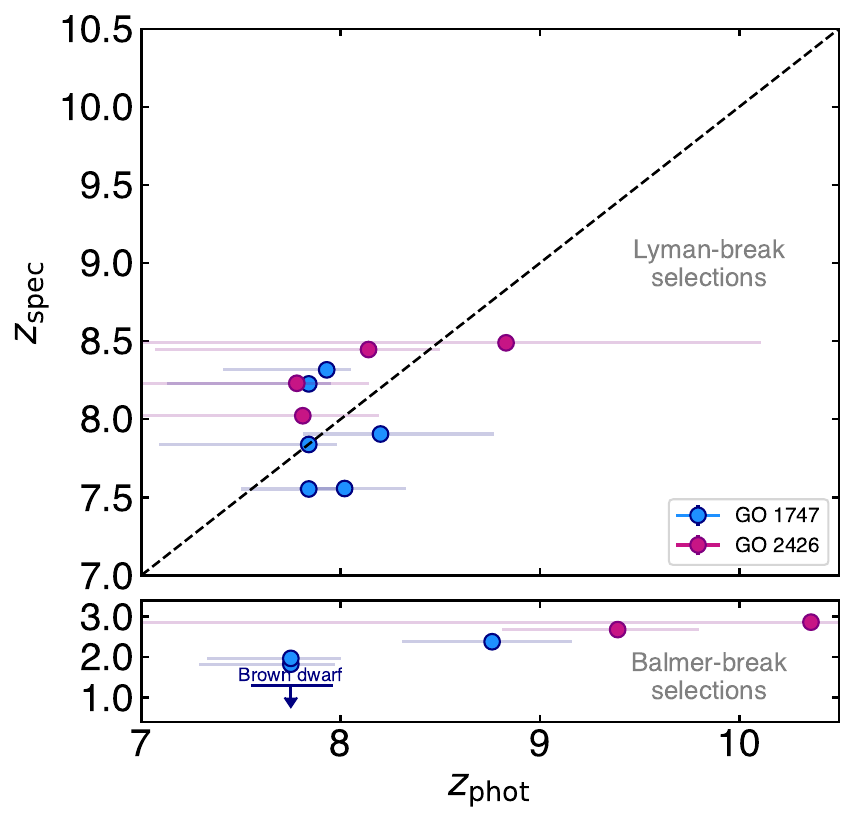}
 \caption{The photometric redshifts from \citet{rb22} (blue circles) and \citet{rojas20} (purple circles) of the primary BoRG candidates observed here, compared to their spectroscopic redshifts as determined from their NIRSpec prism spectra.}
 \label{fig:zspec}
\end{figure}

The primary contaminants in \textit{HST} searches for $z>6$ galaxies are generally brown dwarfs or $z\sim2-4$ galaxies with strong emission lines and/or Balmer breaks. While morphological selections greatly limit the inclusion of brown dwarfs and other point-like sources in high redshift galaxy samples \citep{holwerda14}, the mitigation of low-redshift contaminants is challenging in the absence of the especially deep optical imaging needed to rule out a Balmer break in place of the Lyman-$\alpha$ break. The challenge is illustrated in Figure~\ref{fig:zspec}, where for the primary BoRG targets we compare the photometric and spectroscopic redshifts. We note that the challenge is mitigated with \textit{JWST} selections of dropout samples, since NIRCam provides numerous detection bands redward of the break which can be used to distinguish a Lyman-$\alpha$ break from a Balmer break galaxy.  

For those sources confirmed at high redshift, we find the spectroscopic redshifts are all within or close to the estimated photometric redshifts and their uncertainties, while the galaxy contaminants (described above) all reside at $z\sim2-3$ consistent with the ambiguity between Balmer and Lyman-$\alpha$ break in \textit{HST} dropouts at these redshifts. 

\begin{figure}
\center
\includegraphics[width=\columnwidth]{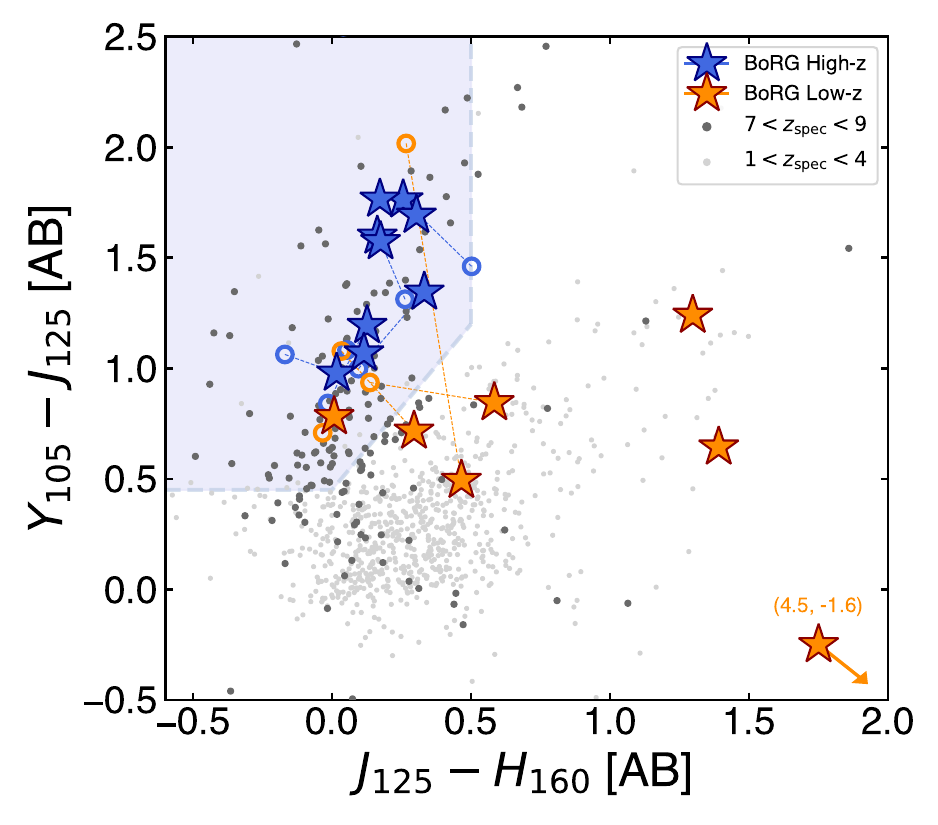}
 \caption{The pseudo-\textit{HST}/WFC3 colors of the primary BoRG-\textit{JWST} sources confirmed in this study, derived using NIRSpec prism spectra and the relevant filter response curves. Sources confirmed at high-$z$ are shown as blue stars, while sources which revealed themselves as low-$z$ contaminants are plotted as orange stars. For those sources originally selected via color-color criteria, we also include their \textit{HST} photometry as empty circles, connected by dashed lines. Lastly, for comparison we also include the NIRSpec spectroscopic compilations of $7<z<9$ sources (dark grey points) and $1<z<4$ sources (light grey points) by \citet{rb24}. The NIR color cuts originally employed for the high-$z$ BoRG selections do an excellent job of segregating the high-redshift population from their lower redshift counterparts. We note that the reddest source indicated by an orange star was expected to be at low redshift as described in footnote~\ref{footnote:par2346}.}
 \label{fig:colors}
\end{figure}

We verify whether the inclusion of those contaminants in the parent samples from which they were selected derives primarily from the choice of color cut, availability of \textit{HST} filters blueward of the Lyman-$\alpha$ break, or other factors. In Figure~\ref{fig:colors} we plot the \textit{spectroscopic} $Y_{\rm 105}-J_{\rm 125}$ and $J_{\rm 125}-H_{\rm 160}$ \textit{HST}/WFC3 colors (i.e., applying the relevant filter response curves to each prism spectrum) of the combined sample of primary targets, to explore whether widely-used color cuts serve as effective measures to exclude low-redshift contaminants from high-$z$ samples. 

We note, however, that only targets from the GO 1747 program were originally selected employing such color cuts, while the targets in GO 2426 were selected primarily using a photometric redshift analysis. For comparison, we also plot the same colors for other NIRSpec-confirmed sources at $7<z<9$ and $1<z<4$, using the data reductions of \citet{rb24}. We find overall the adopted NIR selections perform well in their isolation of $z>7$ galaxies, based on the measured pseudo-\textit{HST} colors. In fact, the primary divide between the high-$z$ and low-$z$ locuses is given by $Y_{\rm 105}-J_{\rm 125} > 1.5\cdot(J_{\rm 125}-H_{\rm 160}) + 4.5$, rather than $Y_{\rm 105}-J_{\rm 125} > 0.45$ or $J_{\rm 125}-H_{\rm 160} < 0.5$ which target large Lyman-breaks and blue UV slopes. While both of the latter have their own merits, \textit{JWST} has shown the high-$z$ population to be far more diverse than previously thought, suggesting both those cuts could appear too restrictive in their high-$z$ selections (e.g., excluding fainter objects or those with redder slopes from mature stellar populations). Although no set of NIR cuts can guarantee 100\% purity, the overlap between high-$z$ and low-$z$ objects from BoRG-\textit{JWST} and the literature appears small, adding credence to their use. We also find no pattern between the availability of blue \textit{HST} filters and the redshift solution of the source -- i.e., most BoRG-\textit{JWST} sources have only a single blue filter (generally F350LP, F600LP, or F606W), the number of which does not appear to be a critical constraint in removing low-$z$ contaminants. 

Of the low-$z$ contaminants confirmed with NIRSpec, three of these were selected solely through photometric redshift estimates (the three contaminants with $J_{\rm 125}-H_{\rm 160}>1$ in Figure~\ref{fig:colors}, from GO 2426), and thus their positions on the color-color diagram is not necessarily surprising. GO 2426 employed a photometric redshift selection with the goal of building as complete a selection as possible, including sources that may lie outside a traditional color window. However, such a selection comes at the expense of higher contamination from lower-redshift galaxies\footnote{We note that the reddest source in Figure~\ref{fig:colors} was removed from the high-redshift sample in \citet{bagley24} based on \spitzer/IRAC photometry, but that this analysis was performed after the target had already been incorporated into the JWST schedule.\label{footnote:par2346}}, as can be seen in Figure~\ref{fig:colors}. Indeed, the contaminants from GO 2426 have a secondary peak in their redshift probability distributions at $z\sim2-3$, and were located close to the selection boundary in integrated redshift probability space. These spectroscopic observations can therefore be used to further inform photometric redshift selections for sources with multiple probability peaks.
The brown dwarf contaminant (orange star within the blue high-$z$ selection of Figure~\ref{fig:colors}, from GO 1747) instead relied on a combination of NIR color cuts, photo-$z$ estimates, and a comparison to the SpeX prism library of $M-$, $L-$ and $T-$dwarf templates of various spectral types, highlighting the challenges of isolating high-$z$ sources from contaminants. 

For the remaining three GO 1747 contaminants residing close to the high-$z$ selection boundary, a comparison to their \textit{photometric} \textit{HST} colors display comparatively brighter $J_{\rm 125}$ fluxes, boosting their colors into the high-$z$ space of the diagram (marked by circles and connecting lines). The reason for such a boost is likely contamination by cosmic rays in one or a number of $J_{\rm 125}$ exposures\footnote{An important limitation of \textit{HST} pure-parallel observations is the absence of dithering, in order to avoid conflicts with the primary observation, which increases the probability of a cosmic ray hit propagating into the final images. Such a limitation will not apply to future pure-parallel \textit{JWST}/NIRCam programs (e.g., BEACON and PANORAMIC) which benefit from dithering, while the longer wavelength coverage to $\sim4.5 \mu$m will greatly aid in the exclusion of low-$z$ interlopers when accounting for their especially red rest-frame optical continua.}. Despite the four examples of contaminants (one brown dwarf and three $z\sim1-3$ galaxies) persisting through NIR color-color selection likely due to cosmic rays, the success rate of the BoRG-\textit{JWST} spectroscopy and a comparison to sources in the literature suggest similar color-color applications over both deep and shallower data sets can prove effective in maximising the purity of high-$z$ samples.

\subsection{Physical Properties of the Most Luminous Sources}
\begin{figure*}
\center
\includegraphics[width=\textwidth]{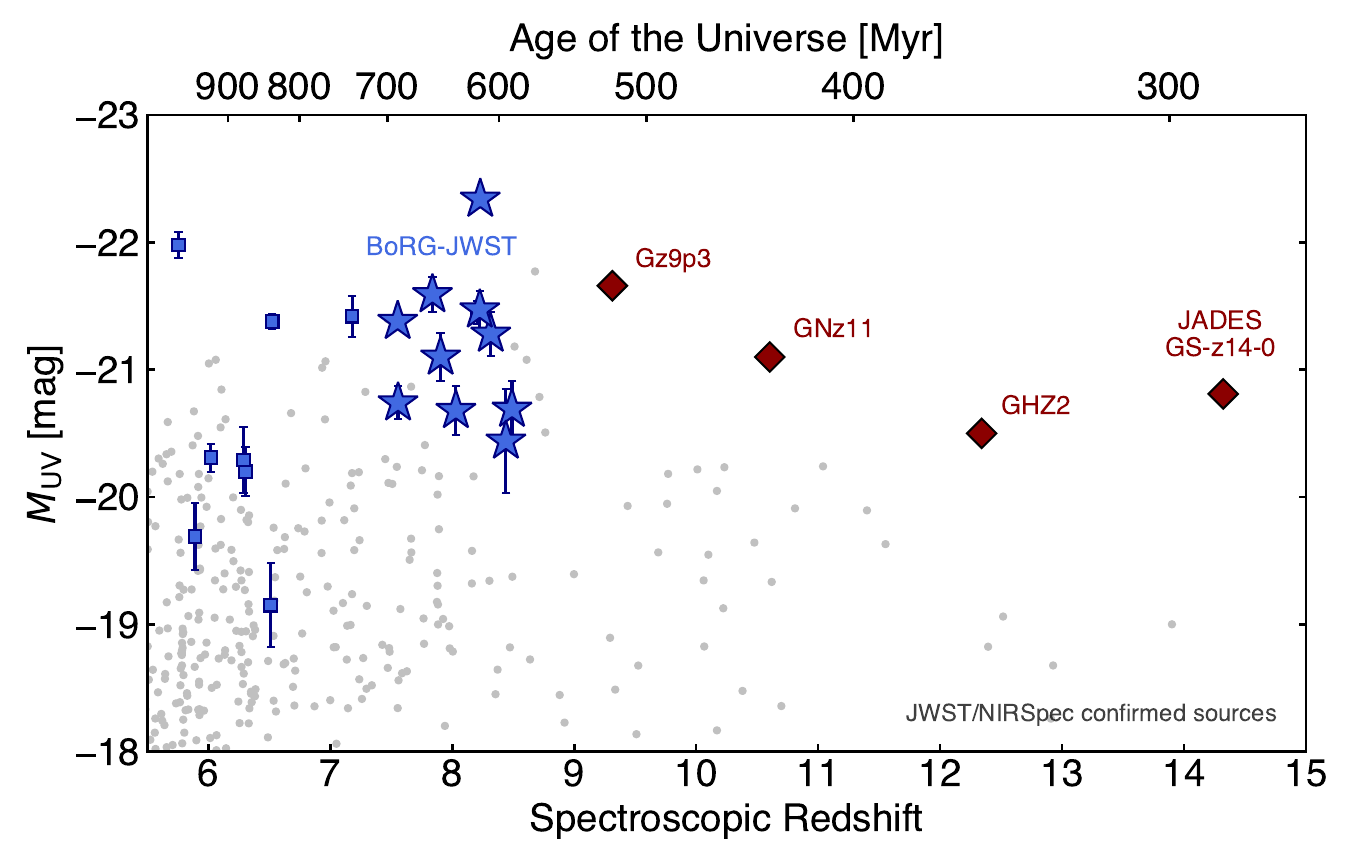}
 \caption{The distribution of galaxy absolute magnitudes and spectroscopic redshifts for the full, high-redshift BoRG-\textit{JWST} sample (blue) and confirmed sources in the literature (grey points from the compilation of \citealt{rb24}). The sources are divided into primary targets (blue stars) and filler targets (blue squares). Red points indicate the most luminous, confirmed sources at $z>9$, including Gz9p3 \citep{boyett23}, GNz11 \citep{bunker23}, GHZ2 \citep{castellano24}, and JADES-GS-z14-0 \citep{carniani24}. The primary BoRG-\textit{JWST} sources represent some of the most luminous confirmed sources within the Epoch of reionization to date.}
 \label{fig:muv}
\end{figure*}

We determine a number of global properties for our confirmed sources, based on their \textit{HST} photometry and NIRSpec spectroscopy, to place them into context with the bulk of confirmed $z>5$ sources in the literature. Given their especially bright apparent magnitudes, we first explore whether any of our sources might benefit from gravitational lensing from a nearby neighbour. We find one target (par0953$+$5153\_1655) has a neighbour only 2.1 arcsec away, which could act as a lensing source. A magnification factor for the background source was calculated following the procedure outlined in and the simulations performed by \citet{rojas20}. Briefly, the gravitational potential of the magnifying source was estimated following the prescription of \citet{mason15}, where the Einstein radius is determined from the separation between the target and the magnifying neighbour. The magnification factor, $\mu$, was then determined from 1,000 Monte Carlo simulations sampling the photometric redshift probability distribution function for both the target and the neighbour while adjusting the $H_{\rm 160}$ flux of the lensing source within its uncertainty. The mean magnification factor was found to be $\mu=2.22$, and we correct all relevant properties in Table~\ref{tab:confirmations} for par0953$+$5153\_1655 by this factor.
Adopting the spectroscopic redshifts listed in Table~\ref{tab:confirmations}, and the mean UV photometric fluxes redwards of the Lyman-break, we determine absolute magnitudes in the range $-20.44<M_{\rm UV}<-22.41$ mag for the primary targets, or $-19.15<M_{\rm UV}<-22.41$ mag if we extend the measurement to our filler confirmations. The redshifts and absolute magnitudes are plotted in Figure~\ref{fig:muv}, where we compare to the spectroscopic compilation by \citet{rb24} as well as values for the most luminous confirmed-sources at $z>9$ (Gz8p3, GNz11, GHZ2, and the two $z\sim14$ JADES sources, from \citealt{boyett23}, \citealt{bunker23}, \citealt{castellano24}, and \citealt{carniani24}, respectively).

The primary BoRG sample presented here is comparable in absolute magnitude range to those of Gz9p3, GNz11, GHZ2, and JADES-GS-z14-0, and clearly contains some of the most luminous $z>7$ sources confirmed in the literature thus far (see also e.g., \citealt{bouwens21_rebels}). Our observations effectively double the number of \textit{JWST}-confirmed $M_{\rm UV}<-21$ mag sources at $7<z<9$, and one source in particular (par0956$+$2847\_169) represents possibly the most luminous source yet discovered by \textit{JWST} within the first billion years of the Universe,
further highlighting the value of the BoRG-\textit{JWST} survey and pure-parallel searches of luminous galaxies.

\begin{figure}
\center
\includegraphics[width=\columnwidth]{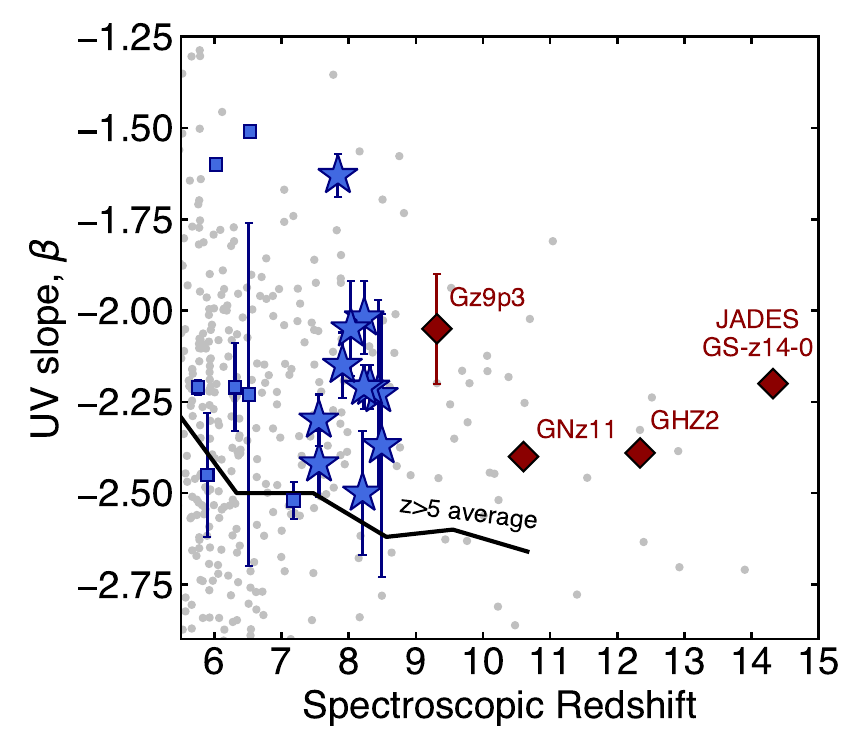}
 \caption{The same as Figure~\ref{fig:muv} but showcasing UV continuum fluxes as a function of redshift, as well as the median values found in stacked spectra by \citet{rb24}.}
 \label{fig:beta}
\end{figure}

We compare the UV slopes of our confirmed sources, based on a power-law fit to the spectrum over wavelengths $\lambda_{\rm rest}=1600-3500$ \AA\ (masking the wavelength range 1850-1950 \AA\ to avoid contamination by possible [\ciii] emission; \citealt{rb24}) and plot the resulting values in Figure~\ref{fig:beta}. We find the BoRG-\textit{JWST} sources display a large range of UV slopes (see Table~\ref{tab:confirmations}), predominantly ranging between $\beta\simeq-2.5$ to $\beta\simeq-2.0$ with evidence for the most luminous sources displaying the reddest slopes likely owing to more evolved stellar and ISM conditions. This is highlighted by the reddest sources in our sample, 1747\_m5, a filler $z\simeq6.5$ source with a clear Balmer break indicative of mature stellar populations.

None of our sources display especially blue slopes below $\beta\lesssim-2.8$ resemblant of particularly metal-poor systems and Lyman continuum leakers, and in general the sample overlaps well with other confirmed luminous sources in the literature (e.g., Gz9p3, GNz11, GHZ2, and JADES-GS-z14-0) while falling well above the median values found from stacked prism spectra \citep{rb24}. Considering the $\sim$2-3 mag fainter magnitudes probed by the latter, as well as the clear trend of redder slopes with enhanced luminosities, the differences in UV slopes can likely be attributed to the luminosity differences between the samples and the range of values is suggestive of some possible dust obscuration.

Lastly, we also verify whether any of our confirmed $z>5$ sources display evidence for Type 1 AGN from broadened Balmer features relative to the widths of forbidden lines. Specifically, we fit the H$\beta$, [\oiii]$\lambda$4960 \AA, and [\oiii]$\lambda$5008 \AA\ lines simultaneously with single Gaussian profiles to verify whether additional components are needed. We find none of our confirmed sources display evidence for additional broad components and thus do not appear to host strong, Type 1 AGN. We caution however that deeper and higher resolution spectroscopy is needed to corroborate these claims, while Type 2 AGN are not considered here due to the inefficiency of standard narrow-line ``BPT'' separations at high redshift. While strong rest-optical lines are identified in individual spectra, we do not identify any especially strong rest-frame UV lines (e.g., \civ$\lambda$1549 \AA, \heii$\lambda$1640 \AA, \oiii]$\lambda$1663 \AA, \ciii]$\lambda$1909 \AA) found in a number of remarkably luminous sources at $z>10$ (e.g., GNz11 and GHZ2; \citealt{bunker23,castellano24}) or Nitrogen-enhanced sources at lower redshift \citep{rui24,topping24,schaerer24}, although this is likely in part due to insufficient signal-to-noise and spectral resolution. Indeed, a UV continuum-normalized stack of all our $z>5$ sources reveals clear \ciii]$\lambda$1909 \AA\ emission comparable to that seen in the bulk of the high-$z$ population with prism observations, in addition to clear rest-optical [\oii]$\lambda\lambda$3727,3729 \AA, [\neiii]$\lambda\lambda$3869,3968 \AA, H$\delta\lambda$4102 \AA, H$\gamma\lambda$4341 \AA\ plus [\oiii]$\lambda$4364 \AA\ (blended), \hei$\lambda$5877 \AA, H$\alpha\lambda$6564 \AA, and [\sii]$\lambda\lambda$6718,6732 \AA\ line emission (c.f. with \citealt{rb24}). A detailed analysis of these sources' metallicities and ionization conditions is left to a future study.

\subsection{Legacy Data Sets for Community Use}
\begin{figure}
\center
\includegraphics[width=0.875\columnwidth]{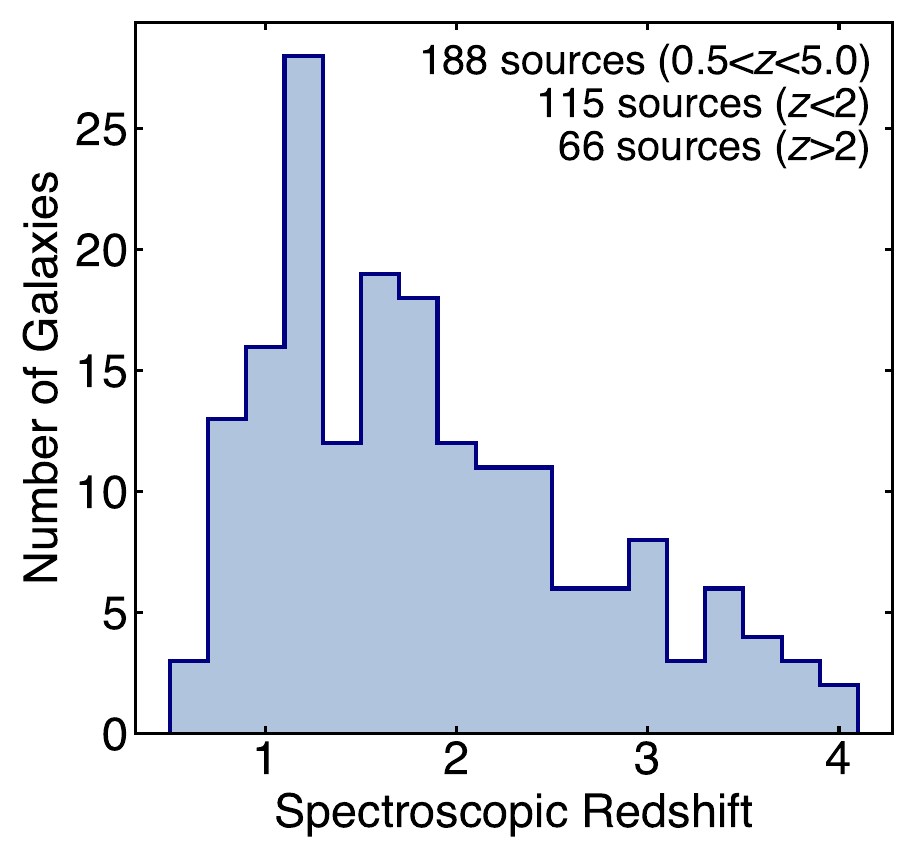}
 \caption{The spectroscopic redshift distribution of $0.8\lesssim z\lesssim4.0$ filler sources included in the GO 1747 MSA setup, for legacy science.}
 \label{fig:fillers}
\end{figure}

As mentioned in Section~\ref{sec:sample}, one of the goals of the GO 1747 program was to make use of the MSA pointings to obtain spectra for a broader selection of galaxies within a range of redshifts. As such, while the highest priority was given to $z>7$ objects, the MSA was also filled with candidate massive galaxies (log\,$M_{*}/M_{\odot}>10.5$) at $1<z<3$, photo-$z$ selected sources at $z\simeq1-6$, and protocluster candidates. In addition to the $z>5$ confirmed in Section~\ref{sec:confirmations}, we confirm an additional 188 galaxies via line emission or continuum breaks and 3 brown dwarfs, highlighting the legacy value of the program for a variety of science goals. The measured redshifts span $0.5\lesssim z\lesssim5.0$ and are plotted in Figure~\ref{fig:fillers} for reference. A full spectroscopic catalog of both the primary and the filler objects will be released on the MAST website for community use.

\section{Summary}
\label{sec:summary}
We present an overview of the BoRG-\textit{JWST} survey, a combination of two \textit{JWST}/NIRSpec Cycle 1 programs (GO 1747 and GO 2426) building on observations taken with the \textit{Hubble} Space Telescope. The program delivers NIRSpec $R\sim30-300$ prism follow up of $7<z<10$ galaxy candidates selected from three independent searches across $>$300 uncorrelated pointings with optical-to-NIR WFC3 imaging from WFC3.

The survey design was centered around the exploitation of the low cosmic variance delivered by the independent sighlines, and the addressing of three primary science questions: (i) How prevalent are $z\sim8-9$ UV-luminous sources? (ii) What is the neutrality of the IGM at $z\sim8$ and which sources carve out early ionized bubbles? (iii) What are the ionizing capabilities of UV-luminous sources compared to their fainter counterparts?

Out of the 18 pointings carried out at the time of writing, we confirm a total of 19 sources at $z>5$, of which 10 represent primary targets. We compare their absolute magnitudes and UV continuum slopes to other spectroscopically-confirmed sources in the literature, finding them among the most luminous sources known to date and consistent with the properties of their peers. While the majority of primary targets revealed themselves as high-$z$ sources, a number (6) also revealed themselves to be either a brown dwarf or $z\sim1-3$ interlopers, and we offer a comparison of both their photometric versus spectroscopic redshifts and \textit{HST} NIR colors in a bid to distinguish the main reason for contamination. 

Lastly, we highlight the legacy value of the BoRG-\textit{JWST} survey by presenting the data release not only of the high-$z$ sample, but also of 191 filler sources observed as secondary targets in the NIRSpec Micro-Shutter Assembly. The combined BoRG-\textit{JWST} sample offers a unbiased, and spectroscopically-confirmed sample, with cosmic variances matched only by the widest \textit{JWST}/NIRCam surveys and upcoming Euclid observations. Such a survey builds on \textit{Hubble}'s legacy and will serve as a unique benchmark for comparisons to observations over deep fields, as well as a template for future pure-parallel searches of high-$z$ sources with \textit{JWST}/NIRCam.

\acknowledgments
This research was funded in whole or in part by the Swiss National Science Foundation (SNSF) [Grant number 210558]. GRB extends his thanks to Beth Perriello and Nimisha Kumari for their invaluable help in coordinating the observations for GO 1747.

This work is based on observations made with the NASA/ESA/CSA James Webb Space Telescope. The data were obtained from the Mikulski Archive for Space Telescopes at the Space Telescope Science Institute, which is operated by the Association of Universities for Research in Astronomy, Inc., under NASA contract NAS 5-03127 for JWST. These observations are associated with programs \# 1747 and 2464. Funding from NASA through STScI awards JWST-GO-01747 and JWST-GO-02426 is gratefully acknowledged.

CM acknowledges support by the VILLUM FONDEN under grant 37459 and the Carlsberg Foundation under grant CF22-1322. The Cosmic Dawn Center (DAWN) is funded by the Danish National Research Foundation under grant DNRF140.

The Flatiron Institute is a division of the Simons Foundation.

\bibliography{sample63}{}
\bibliographystyle{aasjournal}

\clearpage
\appendix
\onecolumngrid

\begin{figure*}[h]
\center
\includegraphics[width=0.975\textwidth]{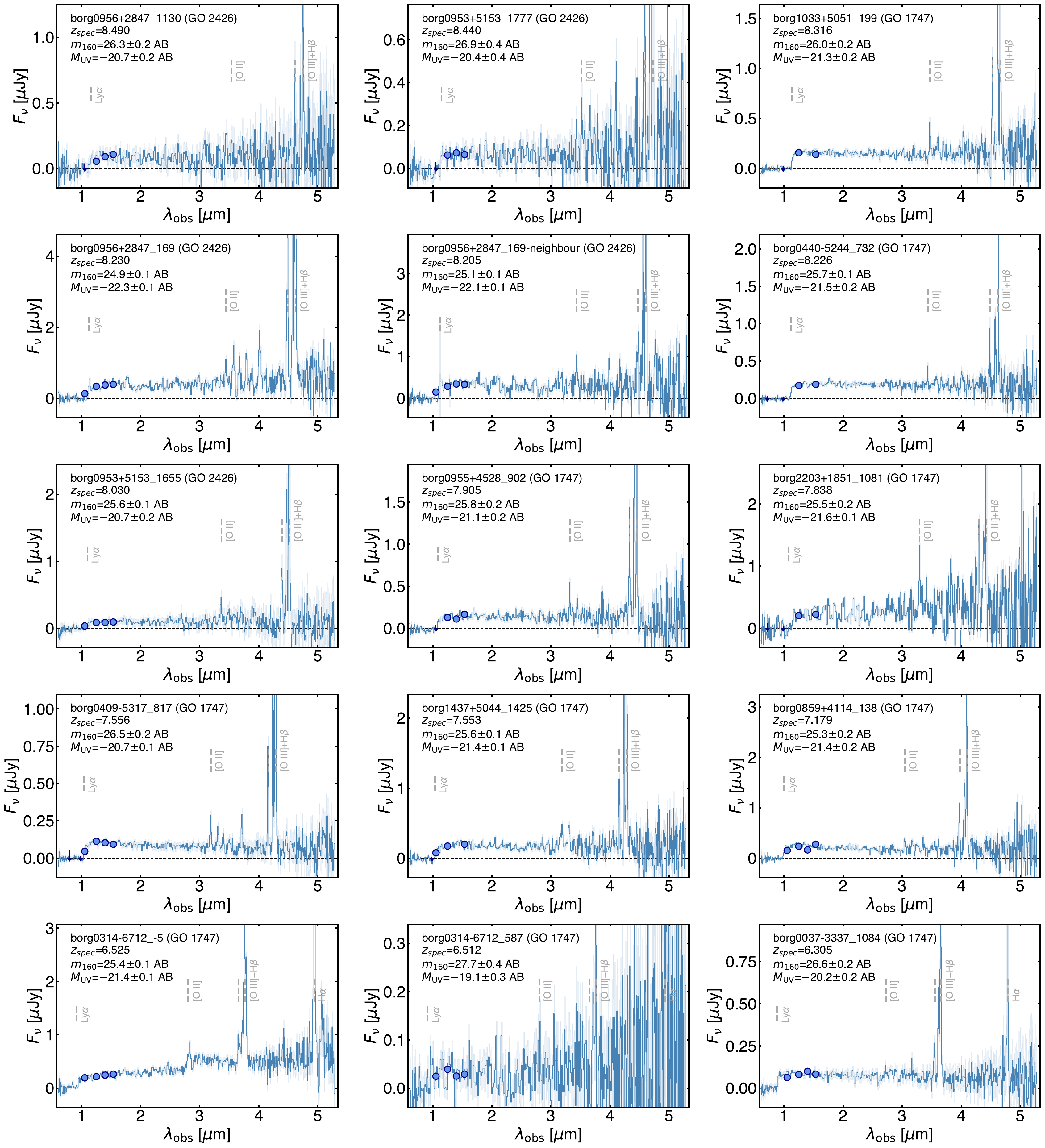}
 \caption{The NIRSpec prism spectra (blue, with grey $1\sigma$ uncertainties) and associated \hst\ photometry (orange circles) of the $z>5$ confirmations from the BoRG-JWST program. The locations of the Lyman-$\alpha$ break and some major emission lines ([\oii]$\lambda\lambda$3727,3730 \AA, H$\beta$, [\oiii]$\lambda\lambda$4960,5008 \AA), regardless of detection or not, are indicated by dashed grey lines and text.}
 \label{fig:spectra}
\end{figure*}

\begin{figure*}
\ContinuedFloat
\center
\includegraphics[width=0.975\textwidth]{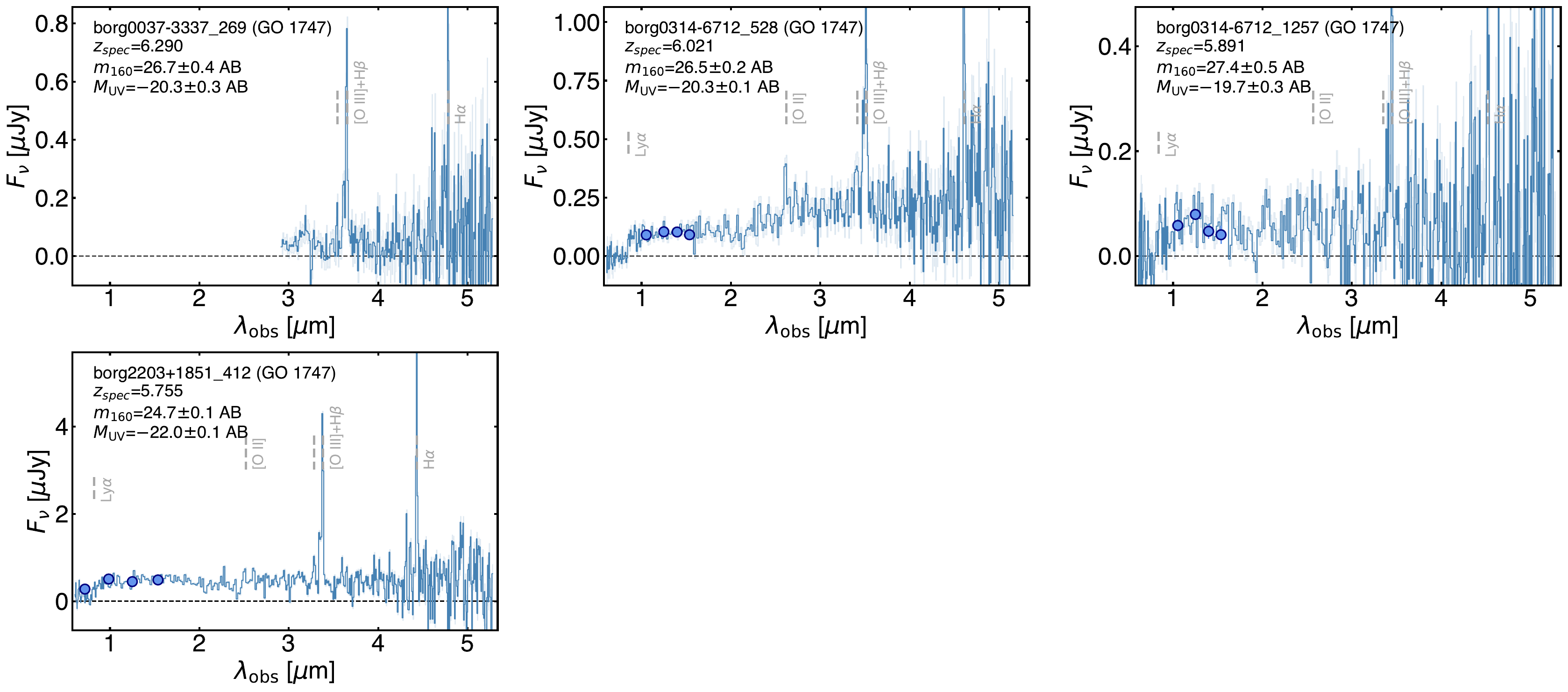}
 \caption{Continued.}
 \label{fig:spectra_continued}
\end{figure*}

\begin{figure*}
\center
\includegraphics[width=0.975\textwidth]{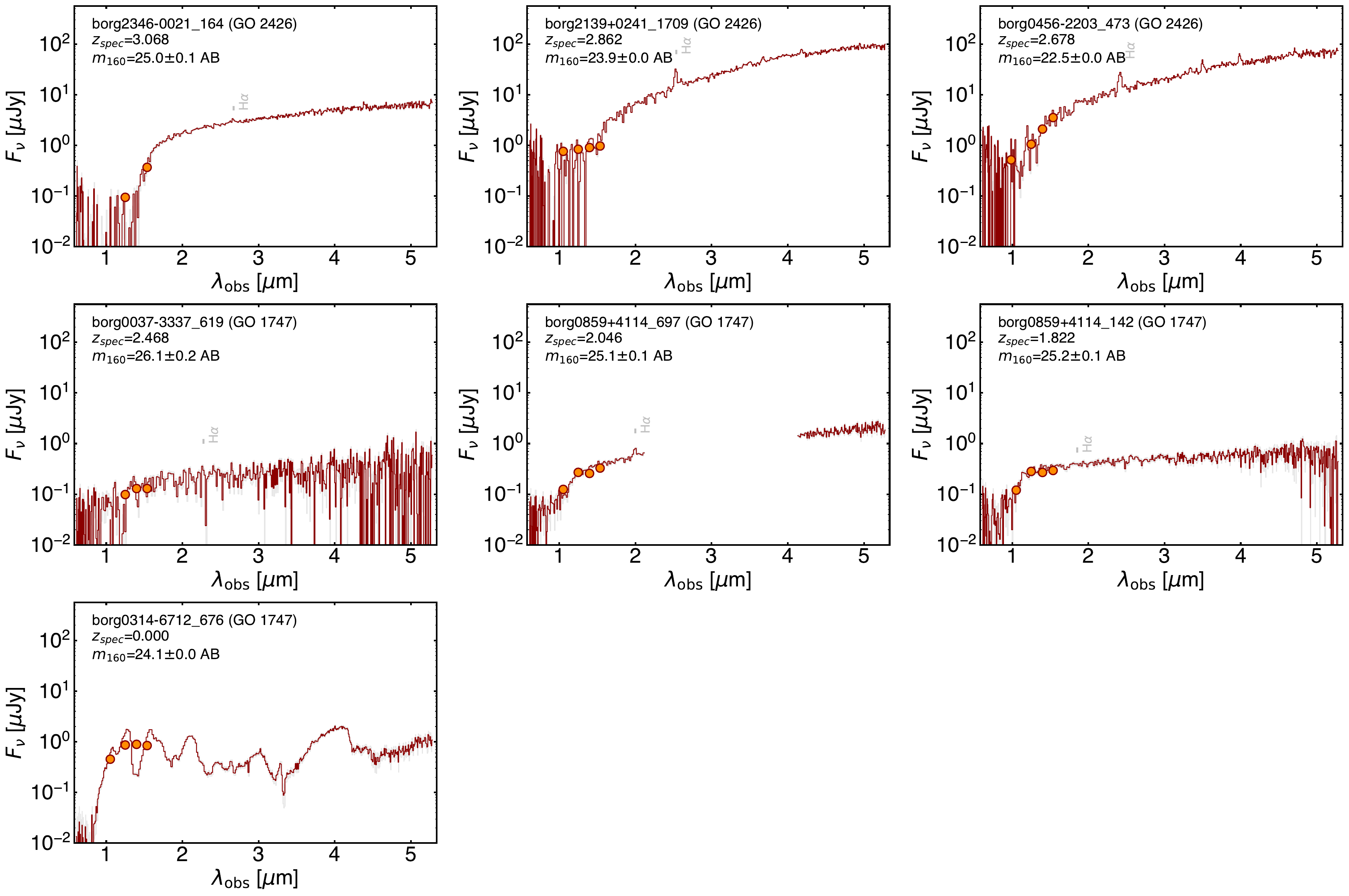}
 \caption{The same as Figure~\ref{fig:spectra} but for the primary BoRG objects which revealed themselves as low-redshift interlopers.}
 \label{fig:lowz}
\end{figure*}

\end{document}